\documentclass{osa-article}

\journal{oe}
\usepackage{bm}
\usepackage{siunitx}
\usepackage[normalem]{ulem}

\articletype{Research Article}

\begin{document}

\title{Numerical modeling of optical modes in topological soft matter}

\author{Urban Mur,\authormark{1,*} Miha Ravnik,\authormark{1,2}}

\address{\authormark{1}Faculty of Mathematics and Physics, University of Ljubljana, Ljubljana, Slovenia \\
\authormark{2}Jožef Stefan Institute, Ljubljana, Slovenia}

\email{\authormark{*}urban.mur@fmf.uni-lj.si} 


\begin{abstract}
Vector and vortex laser beams are desired in many applications and are usually created by manipulating the laser output or by inserting optical components in the laser cavity. Distinctly, inserting liquid crystals into the laser cavity allows for extensive control over the emitted light due to their high susceptibility to external fields and birefringent nature. In this work we demonstrate diverse optical modes for lasing as enabled and stablised by topological birefringent soft matter structures using numerical modelling. We show diverse structuring of light---with different 3D intensity and polarization profiles---as realised by topological soft matter structures in radial nematic droplet, in 2D nematic cavities of different geometry and including topological defects with different charges and winding numbers, in arbitrary varying birefringence fields with topological defects and in pixelated birefringent profiles.  We use custom written FDFD code to calculate emergent electromagnetic eigenmodes. Control over lasing is of a particular interest aiming towards the creation of general intensity, polarization and topologically shaped laser beams. 
\end{abstract}


\section{Introduction}

Light beams can be represented by intensity, polarization and phase \cite{born2013principles, hecht2014optics, zhan2009cylindrical}. Beams with spatially varying polarization are called vector beams and often include polarization singularities \cite{senthilkumaran2020phase}, whereas beams with phase singularities are called vortex beams and carry orbital angular momentum \cite{allen1992orbital, senthilkumaran2020phase}. Both can be used in a diverse array of applications \cite{padgett2017orbital, zhan2009cylindrical} including optical tweezing \cite{padgett2011tweezers}, particle trapping and manipulation \cite{zhan2004trapping}, microscopy \cite{furhapter2005spiral}, imaging \cite{maurer2011spatial}, material processing \cite{niziev1999influence} and optical communications \cite{willner2015optical}.

Typically, laser cavities are designed to generate simple Gaussian beams and then the output beam is manipulated and transformed  \cite{shen2019optical,wang2018recent} to create vortex and vector beams. However, vector beams can be created already within the laser cavity by inserting various optical components, including birefringent components \cite{yonezawa2006generation} and liquid crystals \cite{yoshida2010alignment}. Vortex beams can be created directly within the laser cavity by breaking the orbital angular momentum (OAM) degeneracy of a standard laser cavity, which requires a very specific setup with a series of optical elements \cite{Naidoo2016}. An ideal laser cavity would be geometrically simple and easy to manufacture, but optically capable  of generating arbitrary---vector or vortex---beams, which moreover, would be tunable and switchable. Creating such cavities is an open challenge. 

Due to their unique properties, such as birefringence, self-assembly characteristics and ability to easily tune their optical properties by applying external fields \cite{slussarenko2011tunable, humar2009electrically, du2004electrically}, liquid crystals are employed in a variety of photonic devices. Configuration of liquid crystals is characterized by a nematic director, a headless vector with $\mathbf{n}$ and $\mathbf{-n}$ fully equivalent, which points in the direction of average molecular orientation. The alignment of molecules is reflected in coupling of the material with electromagnetic fields as the direction of optical axis corresponds to the director. Liquid crystals can be used to generate vortex beams by passing light through a q-plate \cite{LcvortexREVIEW}, a radial nematic liquid crystal droplet\cite{Brasselet2009} or through spontaneously formed topological defects\cite{Loussert2013Brasselet} and vector beams by passing a light beam along the disclination line \cite{vcanvcula2014generation}. Light-matter interaction can be bidirectional, with complex beams also influencing the liquid crystal material by optomechanical effects\cite{donato2014polarization}, generation of liquid crystal structures\cite{Smalyukh2010} and induction of liquid-crystal vortices which consequently create optical vortices\cite{Barboza2012}. Liquid crystal based lasers have been demonstrated in different liquid crystal phases \cite{Coles2010} and different geometries\cite{humar2011surfactant}, including Fabry–Pérot cavities \cite{Nys2014,papivc2021topological}. Most studies on liquid crystal lasers however focus on the tunability of the lasing wavelength and not the shaping of the laser beam. Distinctly, in our recent work in \cite{papivc2021topological}, the focus was on how to achieve lasing of selected passive modes and tune them in selected liquid crystal structures.

When the director field is forced in a configuration that cannot be continuously transformed into a uniform state, topological defects appear. Rather recently, different structures of highly nonuniform nematic director were realised that include colloidal crystal \cite{vcopar2015knot}, surface imposed defect structure \cite{berteloot2020ring}, nematic drops with handles \cite{tasinkevych2014splitting} and even knotted and linked nematic fields \cite{machon2014knotted, martinez2014mutually}. Moreover, if chiral nematic (cholesteric) liquid crystal is used in addition to point and linear defect structures in droplets \cite{sevc2012geometrical, posnjak2016points}, other non-trivial and highly twisted configurations including topological solitons \cite{cryst12010094, hess2020control} like torons \cite{varanytsia2017topology}, hopfions \cite{ackerman2017diversity} and skyrmions \cite{nych2017spontaneous} can be achieved in planar cells. Structures can be further stabilized or tailored by use of patterned surface anchoring, metasurfaces and two-photon polymerization \cite{nys2020patterned, tartan2017generation, he2019novel}.  In highly twisted nematic fields and especially in the vicinity of topological defects nematic director field is highly distorted, leading to rapid changes in optical axis direction. Moreover, since electromagnetic fields are polar and nematic director is a non-polar vector, their topological constraints are in principle different, therefore interaction between light and material becomes highly non-trivial. By inserting such topological structures into a lasing cavity, generation of complex modes is expected.

In this paper we numerically calculate diverse complex optical vector modes from generalised lasing cavity based on self-assembled liquid crystal topological structures in a surrounding Fabry–Pérot cavity. We propose a few cavity designs for lasing applications based on liquid crystal structures inside the cavity. We show that well known cylindrical vector modes in a birefringent medium decouple into modes with polarization governed by nematic director field and that different topologies can be patterned into electric field. Theoretical description of liquid crystal polarization alignment to polarization projection is provided. Finally, we give few ideas for creating complex vector modes. More generally, the proposed design based on soft laser cavities has no principal limitation for realisation of arbitrary lasing beam of any intentsity or polarization profile.

\section{Methods}

All the numerical results presented in this article were calculated by a custom written Finite Difference Frequency Domain (FDFD) code, capable of solving a system with arbitrary profile of materials' dielectric permittivity tensor $\bm{\varepsilon}$ in the form of an eigenproblem. The Master equation for magnetic field was formulated from Maxwell curl equations:
\begin{equation}
\mu^{-1}\nabla \times \bm{\varepsilon}^{-1} \nabla \times \mathbf{H} = \omega^2 \mathbf{H},
\end{equation}
where the eigenvalue represents the frequency $\omega$ of the optical mode. Electric field $\mathbf{E}$ in every point of the cavity was calculated from the (nodal) eigenvector $\mathbf{H}$. 

Simulations ran on Intel Xeon nodes with 190GB RAM in Matlab R2019a environment. A fixed number of eigenmodes around the desired wavelength $\lambda$ was calculated in each run, where $\lambda$ in the material was kept large enough to obey the standard rule of thumb determining the voxel size of $\lambda/10$ or smaller. Perfectly matched layers (PML) with the thickness larger than $\lambda/2$ were used to truncate the domain and simulate infinite boundary conditions in transverse directions. Perfect electric conductor (PEC) boundary conditions were used on the top and bottom boundaries to reproduce the effects of perfect mirrors. Applying these symmetries does not affect any results and conclusions, and is used for the only reason of reducing the computational complexity of the problem (especially the needed computer memory). We also performed selected calculations in full geometries (i.e.~without applied symmetries) and identical solutions were obtained. Quality factors were calculated directly from the real and imaginary part of $\omega$ and served as a tool to extract localized modes and classify them. Note, that the presented methodological approach does not include any formulation of gain materials for lasing, thus returning passive resonant modes.  Nevertheless, using same methodological approach, good agreement between -in principle- passive resonant modes and experimentally observed lasing modes is observed in such topological soft matter structures \cite{papivc2021topological}.

Dielectric profile of the cavity was calculated from the nematic order parameter tensor $\bm{Q}$:
\begin{equation}
    Q_{ij} = \frac{S}{2}\left(3n_in_j-\delta_{ij}\right)+\frac{P}{2}\left(e_i^{(1)}e_j^{(1)}-e_i^{(2)}e_j^{(2)}\right)
    \label{orderQ}
\end{equation}
where $S$ is the degree of order and $n_i$ are components of the nematic director. $\mathbf{e}^{(1)}\perp\mathbf{n}$ is the secondary director and $\mathbf{e}^{(2)}=\mathbf{n}\times\mathbf{e}^{(1)}$. The second term in Eq.~\eqref{orderQ} accounts for biaxiality $P$, which quantifies fluctuations around the secondary director $\mathbf{e}^{(1)}$. The anisotropic dielectric tensor $\bm{\varepsilon}$ was calculated from $\bm{Q}$ as \cite{deGennesPG_1993}:
\begin{equation}
\bm{\varepsilon}=\Bar{\varepsilon}\bm{I}+\frac{2}{3}\varepsilon_{\mathrm{a}}^{\mathrm{mol}}\bm{Q},
\label{epsq}
\end{equation}
where $\Bar{\varepsilon}$ is the average dielectric permittivity and $\varepsilon_{\mathrm{a}}^{\mathrm{mol}} = (\varepsilon_{\parallel}-\varepsilon_{\perp})/S$ is the molecular dielectric anisotropy for a degree of order $S$. Nematic director field, which directly affects the dielectric tensor, was determined analytically for topological defects with different topological charges in 3D and winding numbers in 2D, following Refs.~\cite{kleman2006topological,vcanvcula2014generation}.

Nematic soft matter structures have the distinct capability to provide topologically non-trivial birefringent patterns, that affect and possibly  control the polarization, intensity and even angular momentum of the light, either under transmission, in resonators or even in laser cavities \cite{papivc2021topological, Brasselet2009, cardano2012polarization, yoshida2010alignment}. In this work, the non-trivial topology of the nematic fields is imposed by the presence of topological defects, i.e. the $+1$ radial point defects in the radial nematic droplet, and different-winding number disclination lines that span across the nematic layer. These nematic defects strongly affect the topology of the beam polarization profiles---the 2D topological winding number of the polarization in the real-space--- but in given structures \emph{not} the angular momentum of the beams (which is zero), as the cavity does not distinguish between the OAM modes with the opposite handedness due to their identical spatial intensity distributions, radii of curvature on the wavefront and Gouy phase shifts \cite{Naidoo2016}. Note, however, that more generally, especially chiral nematic structures could potentially give rise also to vortex beams with non-zero angular momentum momentum \cite{ackerman2012optical, barboza2013harnessing}.    

\section{Modes in isotropic and radial nematic droplet}

   \begin{figure}[ht]
   \begin{center}
   \begin{tabular}{c}
   \includegraphics[width=\textwidth]{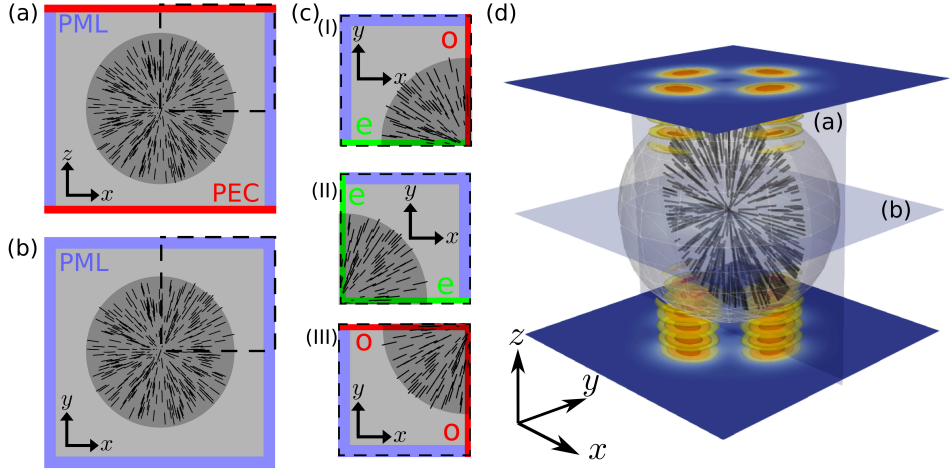}
   \end{tabular}
   \end{center}
   \caption[fig1] 
   { \label{fig:fig1} Simulation setup. (a) Cross section of the system in the $xz$ plane. PEC boundary conditions in $z$ directions model perfect mirrors and PML in lateral directions infinite boundary conditions. Dashed box indicates the actual simulation box. (b) Cross section of the system in the $xy$ plane. (c) Simulation boxes for (I) one even and one odd, (II) two even, and (III) two odd symmetry boundary conditions for electric field. Electric field in the other parts of the system outside the simulation box are obtained by using symmetry transformations. (d) 3D representation of the system with marked positions of the cross sections shown in (a) and (b). Spherical outline represents the radial nematic droplet, which is suspended in the isotropic surrounding medium. Black bars represent the radial nematic director field cross section in the $xz$ plane. Top and bottom planes show the intensity distribution in the intensity maxima closest to the top and bottom reflective surfaces, respectively. Contours represent the isosurfaces of intensity and are only shown outside of the droplet. }

   \end{figure} 

First, we analyzed eigenmodes of a Fabry–Pérot (FP) resonator containing a radial nematic droplet and compared them with the modes of a FP resonator containing a droplet of an optically isotropic material. The droplet was suspended in an isotropic medium and enclosed by two reflective surfaces, simulated by PEC boundary, as shown in Figure~\ref{fig:fig1}(a,b). To achieve sufficient resolution, simulation box was split into eight octants and only one octant with the size of $55\times 55\times 50$ pixels,which roughly corresponds to $5.5\lambda\times 5.5\lambda\times 5\lambda$, was simulated at a time (scheme shown in Fig.~\ref{fig:fig1}(a)). Three different types of octants with PEC boundary conditions in vertical direction and various combinations of even and odd boundary conditions in lateral directions, shown in Fig.~\ref{fig:fig1}(c), were used to cover all possible symmetries of the cross sections of the resulting modes. In each case, two of the lateral boundary surfaces included PML boundary conditions, representing the outer---infinite---boundary of the lasing cavity. Other two lateral boundary conditions were either both set to PMC/PEC or one to PMC (perfect magnetic conductor) and the other to PEC to enforce even or odd boundary condition for electric field, respectively. Electric field outside the simulation box was obtained by using symmetry transformations.

   \begin{figure}[ht]
   \begin{center}
   \begin{tabular}{c}
   \includegraphics[width=\textwidth]{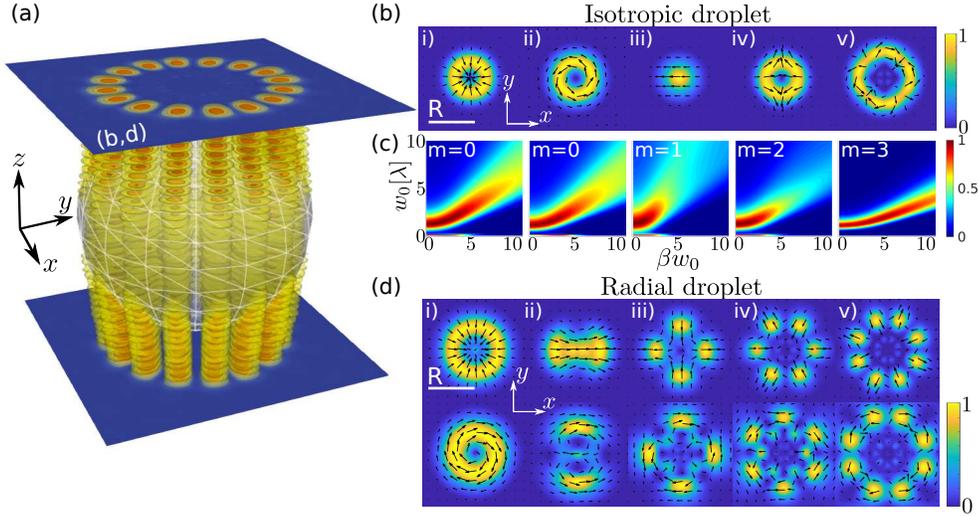}
   \end{tabular}
   \end{center}
   \caption[fig2] 
   { \label{fig:fig2} Modes of isotropic and radial nematic droplet inserted into a FP resonator. (a) 3D intensity distribution of a selected mode, emerging from a radial nematic droplet. (b) Normalized intensity (color code) and polarization (black arrows) cross sections of the eigenmodes in a resonator containing an isotropic droplet. Modes are shown in the plane of the intensity maximum closest to the top reflective surface - the top plane shown in (a). (c) Overlap integrals (color code) between simulated modes emerging from an isotropic droplet and analytical vBG modes with different values of waist size $w_0$ (vertical axis) and $\beta w_0$ (horizontal axis). (d) Normalized intensity (color code) and polarization (black arrows) of the eigenmodes in a resonator containing a radial nematic droplet. Eigenmodes decouple into radially and azimuthally polarized ones in the presence of birefringent material.}
   \end{figure}

Nematic director field with the radial configuration and a $+1$ defect in the middle of the droplet was determined by analytical formula. Radius of the droplet was set to $40$ pixels (roughly $4\lambda$), meaning that the droplet filled $\SI{80}{\%}$ of the FP layer thickness. Ordinary ($n_{\mathrm{o}}$) and extraordinary ($n_{\mathrm{e}}$) refractive indices of liquid crystal and the surrounding isotropic material ($n_{\mathrm{out}}$) were set to $n_{\mathrm{o}} = 1.54$, $n_{\mathrm{e}} = 1.71$, and $n_{\mathrm{out}}=1.47$, respectively, generally corresponding to the indices of widely used 5CB liquid crystal \cite{gray1973new, sharma2021electronic} and glycerol. In the case of the isotropic droplet, its refractive index $n_i$ was set to approximate index of the molten 5CB liquid crystal (i.e.~5CB in the isotropic phase), namely $n_{\mathrm{i}} =\frac{2}{3}n_{\mathrm{o}} + \frac{1}{3}n_{\mathrm{e}} \sim 1.6$.

3D electric field intensity distributions of selected modes are shown in Figure~\ref{fig:fig1}(d) and  Figure \ref{fig:fig2}(a). Optical eigenmodes of a FP resonator containing optically isotropic spherical droplet or particle are cylindrical vector beams \cite{wu2019high} and are shown in Figure \ref{fig:fig2}(b). Following the notation of vector modes in optical fibers, modes in panels (i) and (ii) are known as TE$_{01}$ and TM$_{01}$ modes, while modes in panels (iii) and (iv) are known as hybrid modes---HE$_{11}$ and HE$_{21}$---and contain both, radial and azimuthal electric field contributions \cite{rosales2018review, snyder2012optical}. Mode in panel (v) is one of the higher order hybrid modes, where polarization axis performs multiple turns around the center of the beam.

Mode profiles in Figure \ref{fig:fig2}(b) are shown in the plane close to the top reflecting surface of the cavity, where the refractive index is homogeneous in the lateral direction. Therefore the profile can not be identified as one of the modes that occur in the core region of the step-index optical fibers. In fact, it has been pointed out by Hall in his original article on vector beam solutions of Maxwell's equations \cite{hall1996vector} that the fiber mode would excite a vector Bessel-Gauss free space mode when reaching the end of the fiber---here approximated by a droplet in the center of the cavity---and propagating into the free space (i.e.~homogeneous media). The profiles of the observed simulated vector beams, which are characterized by spatially varying polarization and uniform phase profile, are quantitatively compared to vector Bessel-Gauss (vBG) modes, also referred to as free space solutions in the following, by calculating the overlap integrals between two beam cross sections $\mathbf{E}_{\mathrm{sim}}$ and $\mathbf{E}_{\mathrm{an}}$ as \cite{zaoui2014bridging}:
\begin{equation}
    \eta = \dfrac{|\int \mathbf{E}_{\mathrm{sim}}^* \cdot \mathbf{E}_{\mathrm{an}} \mathrm{d}A|^2 }{\int |\mathbf{E}_\mathrm{sim}|^2 \mathrm{d}A \int |\mathbf{E}_{\mathrm{an}}|^2 \mathrm{d}A},
\end{equation}
where $\mathbf{E}_{\mathrm{sim}}$ is the cross section of numerically simulated mode in the intensity maximum closest to the top reflective surface and $\mathbf{E}_{\mathrm{an}}$ is the cross section of the analytical free space vector beam solution at the distance $z_{\mathrm{max}}$ from the waist of the beam, where $z_{\mathrm{max}}$ equals the distance between the center of the droplet and the plane of the $\mathbf{E}_{\mathrm{sim}}$ cross section.

Overlap integrals were calculated for selected vBG mode numbers $m$ and a range of parameters $w_0\in(0.1\lambda,10\lambda)$ and $\beta w_0 \in (0.1,10)$, by varying both in steps of $0.1$. Resulting overlap integrals for vBG mode number $m$, for which the best agreement is achieved, are shown in Figure \ref{fig:fig2}(c). For all five modes, the agreement is above $\SI{95}{\%}$ ($\SI{99}{\%}$ for (i) and (iii), $\SI{98.5}{\%}$ for (ii), $\SI{98}{\%}$ for (iv) and $\SI{95}{\%}$ for (v)), implicating that vBG beam profiles can in fact be used to describe the modes of the isotropic droplet.

If the birefringent droplet with radial profile of the optical axis/director is inserted in to FP cavity, two families of vector modes with space varying polarizaton profiles are observed: one with polarization purely parallel (radial polarization) and one with polarization purely perpendicular to the director field (azimuthal polarization) of the nematic droplet. Cross sections of mode profiles with the highest Q-factors from both families in the plane of the intensity maximum closest to the upper reflective surface are shown shown in Figure \ref{fig:fig2}(d). In the presence of birefringence the optical modes of the isotropic droplet separate into purely radial and azimuthal contributions. TE$_{01}$ and TM$_{01}$ modes are still found as elementary radial and azimuthal solutions. All the hybrid modes decouple into solutions with multiple intensity maxima and polarization in either radial or azimuthal direction. We label the observed vector beams as VB$^{\mathrm{radial}}_m$ and VB$^{\mathrm{azimuthal}}_m$ modes, where $m$ is the mode number, which is reflected in the $2m$ intensity maxima in the azimuthal direction and is associated with the mode number of the beam emerging from the isotropic droplet (as determined by overlap integrals in Fig.~\ref{fig:fig2}(c)). Phase profiles of the modes are uniform over the cross sections, meaning that the modes do not possess phase singularities (i.e.~optical vortices) and do not carry orbital angular momentum (OAM). Importantly, localization of both polarizations is ensured by refractive index of the surrounding material, which is lower than both, ordinary and extraordinary, indices of liquid crystal.

This discussion shows that the observed vector beams from a laser cavity with an embedded radial nematic droplet show similarities with vBG beams, but up to a varying degree of similarity. Similarly, vector Laguerre-Gaussian \cite{tovar1998production} solutions (i.e.~not the scalar Laguerre-Gauss beams \cite{senthilkumaran2020phase} with the helical phase profile) could be used, however neither of those sets can perfectly describe the emerging modes by only using a single vBG or vLG solution. Therefore, the modes in this work are labeled simply as vector beam modes with the mode numbers $m$ (VB$_m$), meaning that they are in principle unique, as determined by the actual optically anisotropic cavity setup.
   
\section{2D cavities}

\begin{figure}[ht]
   \begin{center}
   \begin{tabular}{c}
   \includegraphics[width=\textwidth]{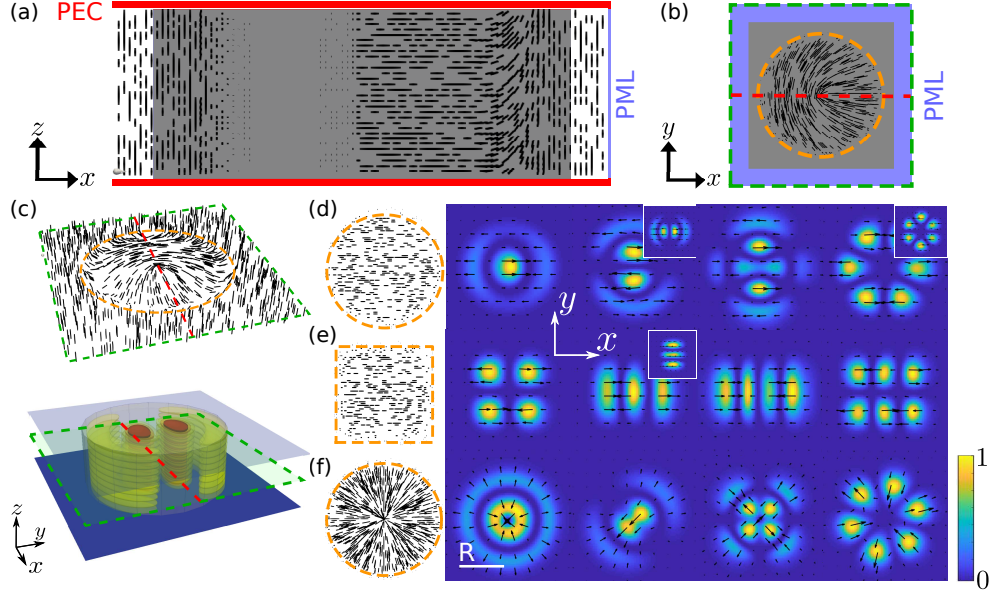}
   \end{tabular}
   \end{center}
   \caption[fig3] 
   { \label{fig:fig3} Design of different birefringent structures for different lasing modes. (a) Side view and (b) top view of the simulation setup. FP resonator contains liquid crystal with the director oriented parallel to the mirrors inside a cylindrical or rectangular region (marked with orange dashed outline) and in the direction normal to the mirrors outside. Projection of the nematic director onto the selected cross section is shown by black bars. (c) 3D representation of director orientation in the plane parallel to the mirrors (marked with green dashed outline) and 3D intensity profile (coloured contours) of a selected mode. Red dashed line in (b) and (c) marks the position of the side view, shown in (a). A selection of eigenmodes emerging from (d) cylindrical region containing homogeneous in-plane director field, (e) rectangular region containing homogeneous in-plane director field and (f) cylindrical region containing a defect line with a $+1$ winding number. In all three cases polarization follows the director field. Insets in (d) and (e) show modes with the same direction of polarization but rotated intensity profiles. Color bars represent normalized electric field intensity. }
   \end{figure} 
   
In order to further study the effects of cavity shape, director field configuration and different topology, we assumed that the Fabry-Perot cavity was filled entirely with liquid crystal. Inside a certain cylindrical or rectangular region the optical axis was directed parallel to the top and bottom mirrors (i.e.~in-plane). Outside the region the director was gradually turned in the direction normal to the mirrors (see Fig.~\ref{fig:fig3}(a)--(c), outline of the cylindrical region is market with orange dashed outline). For simplicity, the director field was invariant in the direction normal to the mirrors, meaning the dielectric profile only varies in two dimensions, hence we name such profiles 2D cavities (although the third dimension is still needed to confine the light). The setup automatically ensured change in refractive index value and localized the modes with extraordinary polarization (i.e.~polarization parallel to the director field), which was previously done by use of surrounding isotropic material, meaning that the cavity shape for the extraordinary polarization was determined by its corresponding refractive index. No symmetry boundary conditions were used. Computational domain had size of $80\times80\times30$ pixels ($\approx 8\lambda\times 8\lambda\times 3\lambda$). Refractive indices were kept the same as in the case of the droplet.

Results for cylindrical and rectangular cavity with the homogeneous in-plane director field are shown in Figure \ref{fig:fig3}(d) and (e). The shape of the intensity profile is clearly directed by the shape of the in-plane director field region, while the direction of the optical axis is directly imprinted in the polarization profile. Insets show that modes with the rotated intensity profile still keep the same direction of polarization.

In Figure \ref{fig:fig3}(f) the in-plane director formed a defect line with a winding number of $+1$ spanning between top and bottom mirrors. Resulting modes are highly similar to the modes, that emerge from a radial nematic droplet, confirming that the polarization is mainly determined by the local direction of in-plane projection of the nematic director. It is noteworthy that the  director manipulation as described here, will only localize modes polarized in the direction of the director field, as the perpendicular polarization sees no refractive index contrast, therefore no azimuthally polarized modes are found.

     \begin{figure}[ht]
   \begin{center}
   \begin{tabular}{c}
   \includegraphics[width=0.96\textwidth]{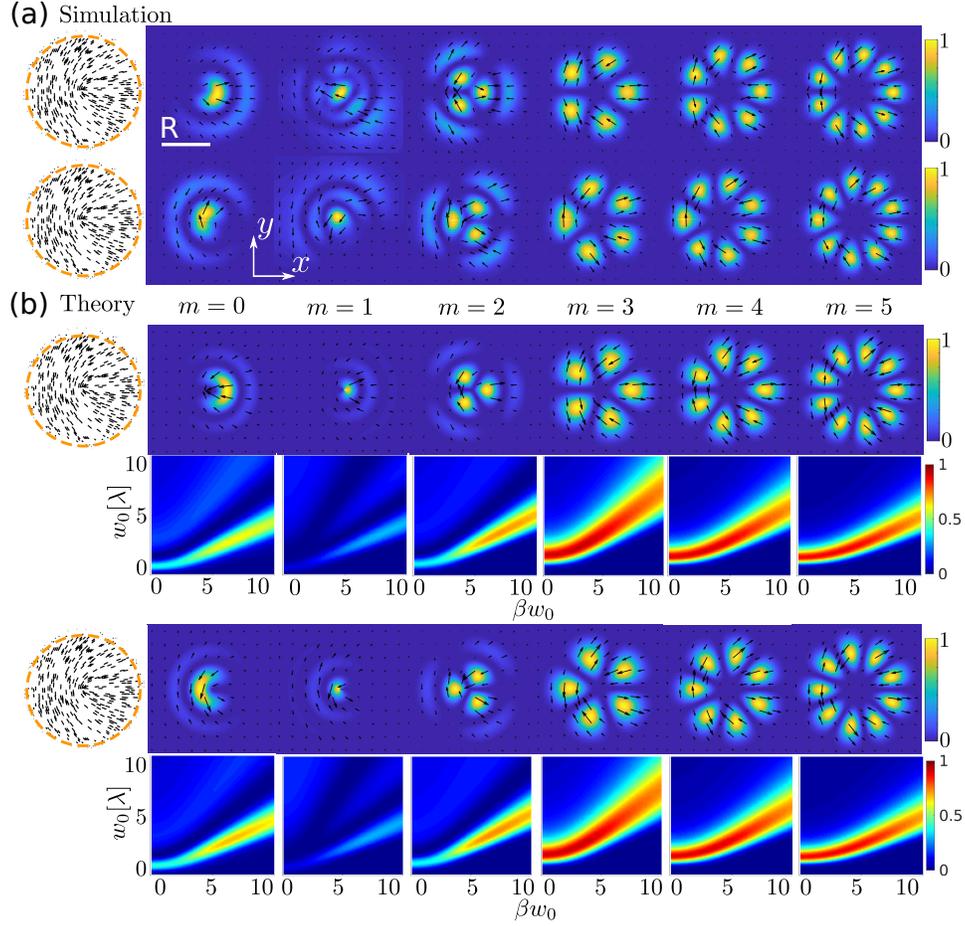}
   \end{tabular}
   \end{center}
   \caption[fig4] 
   { \label{fig:fig4} 
Normalized intensity (color code) and polarization (black arrows) of optical modes emerging from a cylindrical region (marked with orange dashed outline) containing defect lines with half-integer winding numbers. (a) Simulated modes of the cavity containing a defect line with a $+1/2$ winding number. Two families of solutions are found and different polarization profiles. (b) Analytically calculated products of free-space vector beams (here vBG) with mode numbers $m$ and $\mathrm{P}_{1/2}$ polarizer for two different orientations of vBG beam with respect to the polarizer, giving two families of beams. Overlap integrals (color code)  of the free space vBG beams with the actual simulated modes in (a), indicating the parameters $w_0$ and $\beta w_0$ for which the best agreement is achieved.}
   \end{figure} 

In the next step a defect line with winding number of $+1/2$ was studied, a structure which is allowed in a non-polar vector field of nematic director, but not in real vector electric field.  Interestingly, resulting modes once again possess a polarization resembling the underlying director field (Fig.~\ref{fig:fig4}), but with odd number of intensity maxima in the azimuthal direction. Since the direction of the polarization of the emitted modes matches the optical axis direction of the liquid crystal, the liquid crystal acts effectively as a polarizer. Therefore we presuppose that the electric field polarization of the modes emerging around $z$-invariant topological nematic defects $\mathbf{E}_{\mathrm{pol}}$ can be effectively described as a product of free-space solutions of vector Helmholtz equation $\mathbf{E}_{\mathrm{free}}$ (for example Hermite-Gaussian, vector Bessel-Gaussian or vector Laguerre-Gaussian beams) and the polarizer with the optical axis in the direction of the director field around the defect $\mathbf{P}$ as:
\begin{equation}
 \mathbf{E}_{\mathrm{pol}} = (\mathbf{E}_{\mathrm{elem}}\cdot \mathbf{P}) \mathbf{P}.
\label{polarizer}   
\end{equation}
We use analytical beam profiles at $z=0$, since the waist of the emerging beam is in this case expected to be centered at the end of the cylindrical region which acts as a waveguide.

Calculated results for $+1/2$ defect are shown in Fig.~\ref{fig:fig4}, where the $+1/2$ polarizer field is written as $\mathbf{P}_{1/2}(r,\phi) = \cos(-\frac{1}{2}\phi)\mathbf{e}_r + \sin(-\frac{1}{2}\phi)\mathbf{e}_{\phi}$. With the appropriate selection of elementary solutions' parameters for vector Bessel-Gauss beams, good matching with simulations in terms of polarization direction can be achieved. Matching is once again quantified by overlap integrals. For mode numbers $m>2$ matching of up to $\SI{90}{\%}$ is achieved. For smaller mode numbers, matching of about $\SI{30}{\%}$--$\SI{70}{\%}$ is found. The origin of lower matching likely lies in emergence of additional intensity maxima in the lateral direction (1 in the case of $m=0$ and $m=2$, 2 in the case of $m=1$), which cannot be described well by the vector beams, which are solutions of the Helmholtz equation in the free space, while we deal with the spatially varying refractive index. Maxima in the lateral direction emerge due to the fact that vector beams with low mode numbers have smaller radii (as seen from Fig.~\ref{fig:fig2}(b)) and do not fit into the cylindrical region for a wavelength, determined by the distance between mirrors. Modes with low mode numbers and single maximum in the lateral direction are expected to be found at a different ratio between the radius of the cylindrical region and the distance between mirrors or different wavelength. To generalize, local direction of spatially varying optical axis in $z$-invariant birefringent cavity determines the polarization of the emerging mode and can be well described by using free space solutions and Equation \eqref{polarizer}. On the other hand the shape of its intensity profile is determined by the shape of the cavity in terms of refractive index and matching with free space solutions is lower.  

  \section{Constructing complex modes}
  
  \begin{figure}
   \begin{center}
   \begin{tabular}{c} 
   \includegraphics[width=\textwidth]{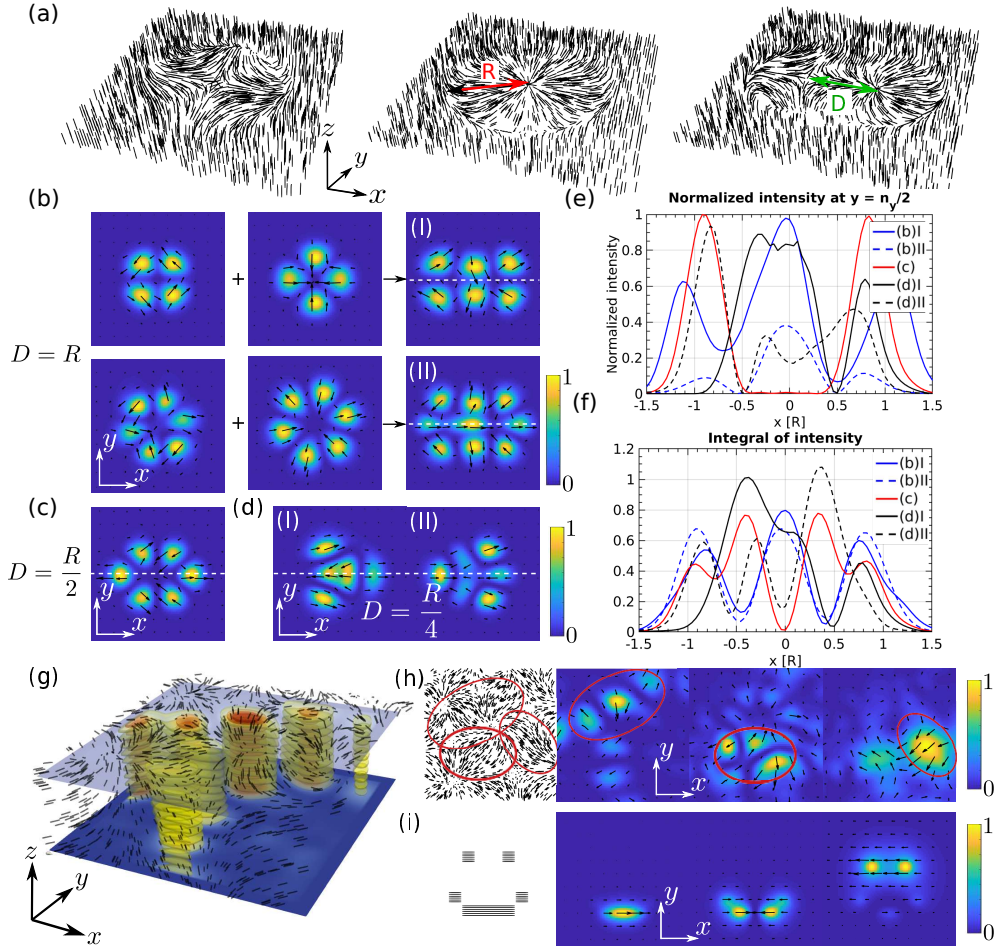}
   \end{tabular}
   \end{center}
   \caption[fig5] 
   { \label{fig:fig5} Possible realizations of more complex intensity and polarization profiles. When cylindrical regions containing a $\pm1$ defect lines (shown in (a)) are brought together, their belonging modes merge into a combined one. (b) Merging of the modes at $D/R=1$. (c) Modes at $D/R=1/2$. Features of isolated mode can hardly be recognized. Complex polarization profiles are a consequence of nematic director field configuration around two defects. (d) Modes at $D/R=1/4$. 
   (e) Normalized intensity of the modes shown in the panels (b) I, (b) II, (c), (d) I and (d) II in the middle of the simulation box in the $y$ direction (marked with white dashed line). (f) Integral of the intensity of the same modes in the $y$ direction. Values are normalized so that the integral over the entire region has a value of 1. (g) 3D intensity distribution and a director field profile of a selected mode from a continuous director field where the director periodically changes its orientation. (h) Modes from  a continuous field are localized in the regions where the director is parallel. Polarization is either parallel or perpendicular to the director field. (i) Optical modes of a pixelated configuration with director field in individual pixels laying either parallel to the mirrors (in-plane director, here shown with black lines) or in the direction perpendicular to the mirrors (out-of-plane director, white regions). Obtained intensity profiles reflect the in-plane director.}
   \end{figure}

    Above presented principles provide an interesting platform for custom design of modes with more complex intensity and polarization profiles. Figure \ref{fig:fig5}(a)--(f) shows that bringing two cylindrical regions with $\pm1$ defect lines in vicinity of each other makes their modes merge into a mode with non-trivial polarization winding. The distance between the defects at which the transition occurs is here mainly governed by the size of cylindrical region, which determines the size of the mode. When the ratio $D/R<2$, regions begin to penetrate into each other and modes start to merge. However until $D/R$ reaches value of 1 features of isolated modes are still visible (Fig.~\ref{fig:fig5}(b)). When $D/R \approx 1$ the cylindrical regions are effectively merged into one, which is reflected in the shape of corresponding modes. Intensity profiles of isolated modes become less prominent. When the $D/R<1$ the shape of the region is becoming more and more cylindrical which is reflected in the shape of the mode, which becomes less elongated and resembles the mode of the isolated defect. However the complex polarization patterns are governed by nematic director field with two topological defects (Fig.~\ref{fig:fig5}(c)). We expect the most interesting effects to occur when the distance between the defects is close to the wavelength of the light (here ~R/4, Fig.~\ref{fig:fig5}(d)). At this point it is hard to recognize modes of isolated defects, shape of the mode is again almost cylindrical, but with a polarization profile which is not observed in the modes of isolated defects. When the distance gets even smaller the wavelength of the light would be too large to be effected by the distortions of the director field. The defects would annihilate when the distance between their cores would get smaller than the correlation length in the LC. To showcase this effects we plot the intensity profile in the cross section at $y=n_y/2$ (marked by white dashed lines in Fig.~\ref{fig:fig5}(b)--(d)), where $n_y$ is the size of the system in $y$ direction (Fig.~\ref{fig:fig5}(e)) and integral of the intensity in the $y$ direction (Fig.~\ref{fig:fig5}(f)).

  A distinct advantage of using liquid crystals as a medium inside cavity is their tunability and susceptibility to external fields. Figure \ref{fig:fig5}(g)--(h) show that localized modes can be found in a continuous nematic director field. A sinusoidal modulation of angle between the mirror plane normal and director field with different periods in orthogonal lateral directions was used. As shown in the first panel, such configuration generates areas of in-plane director field with different profiles. Once again, modes in those regions occur and possess polarization which is parallel or perpendicular to the director field. Director configuration and consequently polarization of emerging modes could be further tailored by use of external fields, patterned surfaces or even by insertion of colloidal particles. Moreover, areas of in- and out-of-plane director field can be also achieved in a pixelated system where the direction is decided separately in each individual pixel with its size comparable to $\lambda$. As shown in Figure \ref{fig:fig5}(i)  intensity profile of the beam can be, in principle, shaped into custom patterns, corresponding to pixels with in-plane director. The wavelengths of all three modes shown differ by less than $\SI{3}{\%}$, size of a pixel is approximately $0.7\lambda\times0.7\lambda$.

\section{Discussion}

In this work, the topology of the resonator-embedded liquid crystal structures combined with the actual geometry of the structures (droplet, layer,...) within the resonator is transferred distinctly into the topology of the light polarization. We extend the work of Ref.~\cite{papivc2021topological}, which recently reported on experimental realization of lasing from liquid crystal structures inserted into Fabry–Pérot (FP) cavity, and here, explain the basic concepts how to design birefringent cavities and their belonging passive modes, distinctly focusing on the role of birefringence in the origin of petal-like modes, their relation to modes of isotropic spheres in FP cavities reported by Ref.~\cite{wu2019high} and free space vector beams. Additionally, we study 2D cavities as a simplified system, which allows us to use a wider variety of elementary defects in the nematic director with different winding numbers. Lastly we show--–as a perspective and possible experimental challenge---that the arbitrary mode profiles can be created in different simplified (pixelated) design or as combination of previously presented cavity profiles.

The calculated passive modes can be used to anticipate the type and profiles of the active (lasing) modes which will emerge from a lasing cavity in the presence of gain media. However, the actual lasing mode profile -as above the lasing threshold- will importantly depend also on gain and pump distribution. Therefore, combinations of these parameters together with the geometry effectively determined within the passive modes will select the actual lasing modes, i.e. which will have the lowest lasing thresholds and what will be the mode profile above the lasing threshold.

Importantly, the spatial profiles of the light polarization presented here are shown to carry a nontrivial (i.e.~nonzero) topological charge---near-field winding of the polarization in the real space at the top of the cavity, which would be an emitting facet of an active laser---(but not the orbital angular momentum---OAM) which is distinctly determined and coupled to the topological charge of the \emph{nematic} topological defects embedded in the FP cavity (e.g.~Fig.~4). Especially in the droplet resonators, such winding of the polarization is expected to be observed also in the far field as the calculated modes show similarities with vector Bessel Gauss beams, quantified by calculated overlap integrals, for which the diffraction and propagation solutions in free space are known to preserve the polarisation topology \cite{greene1998properties}. More generally, this shows a two way coupling between the topology of the LC structures and the topology of the emitted light (polarization). However in order to fully understand the propagation and far field of the general modes created though general liquid crystal topological structures, the actual lasing modes would need to be calculated first and then propagated to the far field regime.

The general approach of soft matter photonics has selected experimental chalenges which one could experience when realizing liquid crystal FP lasers, like photobleaching of dies, alignment of emission and absorption spectra of the materials with the modal spectrum of the cavity, width of the emission spectra of die, compared to spectral distance between the modes (mode selection), dependence of Q-factors and consequently stability of the belonging modes on the used materials' parameters and their birefringence etc. A range of these challenges were mitigated and/or solved in different setups, with examples of topological soft matter applications including for example cholesteric band edge lasers \cite{Coles2010}, blue phase lasers \cite{cao2002lasing}, high Q-factor WGM resonators \cite{humar2009electrically} and others \cite{mysliwiec2021liquid}.

In broader terms the soft matter photonics has gained significant interest in recent years as it can offer alternatives to address some limitations of conventional solid state photonics \cite{kolle2018progress}. Compared to their solid state counterparts, the soft photonic elements are well reconfigurable as they are able to adapt and respond to different changes in the surrounding environment \cite{kolle2018progress}. In liquid crystal systems, the nematic birefringent profile which determines the optical properties can be controlled by applying external electric, magnetic or light fields, by confining the material into cavities and enforcing different surface (anchoring) conditions. Soft photonic elements also can be biocompatible \cite{wu2016optical, humar2017toward} and can self assemble into 3D \cite{kim2011self} and hierarchical structures \cite{zhuo2019hierarchical}, even with very fine feature sizes and high surface smoothness \cite{humar2009electrically}, which makes them useful for plasmonic elements \cite{fan2010self}, metamaterials \cite{chen2020soft} and laser cavities \cite{humar2015intracellular}. Some limitations of the soft matter approach to structuring light include photobleaching \cite{demchenko2020photobleaching}, possible material instability due to flows and evaporation, less control on smaller scales, and  complex self-assembly mechanisms \cite{van2018grand}. 

\section{Conclusion}

In conclusion, we have demonstrated that birefringent nematic structures inserted into Fabry–Pérot cavity highly impact the optical modes compared to the modes that would emerge from similar cavity based on isotropic components, especially, by decoupling the modes into families with predominant perpendicular or parallel polarization. Also, we show that the nematic director field configuration gets directly patterned in the modal electric field polarization, including through different topologies and cavity shapes, as determined by the (possibly polarization dependent) effective refractive index profiles. We have also shown that complex laser vector beam profiles can be constructed by use of space variant nematic fields, pixelated systems and topological defects, enabling the possibility to tune the light beam shape and polarization by external stimuli. 

More generally, the work provides an insight into the control of the intensity and polarization of the passive light modes by using topological soft matter structures based on birefringent profiles of liquid crystals inserted into the Fabry–Pérot cavities. In addition to studies of active modes and lasing thresholds, an interesting extended approach would also be to use more complex---possibly chiral---liquid crystals structures, such as torons and cholesteric droplets to perform either as resonators or laser cavities, possibly also leading to structured light with topological helical light modality with net nonzero angular momentum. 
Soft photonic elements could also perform as building blocks of photonic chips \cite{zhang2015organic, lu2017soft} or light-by-light controllable logical devices \cite{fu2012all,salmanpour2015photonic}.

\section*{Funding}
This research was supported by the Slovenian Office of Science (ARRS) through grants J1-1697, P1-0099, N1-0195 and EU Horizon 2020 ERC advanced grant LOGOS.

\section*{Acknowledgments}
The authors acknowledge M.~Papič, M.~Humar and I.~Muševič for valuable experimental input on the explored topic.

\section*{Disclosures}
The authors declare no conflicts of interest.

\section*{Data availability}
 Data underlying the results presented in this paper are not publicly available at this time but may be obtained from the authors upon reasonable request.

\bibliography{main}

\begin{thebibliography}{10}
\newcommand{\enquote}[1]{``#1''}

\bibitem{born2013principles}
M.~Born and E.~Wolf, \emph{Principles of optics: electromagnetic theory of
  propagation, interference and diffraction of light} (Elsevier, 2013).

\bibitem{hecht2014optics}
E.~Hecht, \emph{Optics}, Always learning (Pearson, 2014).

\bibitem{zhan2009cylindrical}
Q.~Zhan, \enquote{Cylindrical vector beams: from mathematical concepts to
  applications,} {\protect\JournalTitle{Advances in Optics and Photonics}}
  \textbf{1}, 1--57 (2009).

\bibitem{senthilkumaran2020phase}
P.~Senthilkumaran and S.~K. Pal, \enquote{Phase singularities to polarization
  singularities,} {\protect\JournalTitle{International Journal of Optics}}
  \textbf{2020}, 2812803 (2020).

\bibitem{allen1992orbital}
L.~Allen, M.~W. Beijersbergen, R.~Spreeuw, and J.~Woerdman, \enquote{Orbital
  angular momentum of light and the transformation of laguerre-gaussian laser
  modes,} {\protect\JournalTitle{Physical review A}} \textbf{45}, 8185 (1992).

\bibitem{padgett2017orbital}
M.~J. Padgett, \enquote{Orbital angular momentum 25 years on,}
  {\protect\JournalTitle{Optics express}} \textbf{25}, 11265--11274 (2017).

\bibitem{padgett2011tweezers}
M.~Padgett and R.~Bowman, \enquote{Tweezers with a twist,}
  {\protect\JournalTitle{Nature photonics}} \textbf{5}, 343--348 (2011).

\bibitem{zhan2004trapping}
Q.~Zhan, \enquote{Trapping metallic rayleigh particles with radial
  polarization,} {\protect\JournalTitle{Optics express}} \textbf{12},
  3377--3382 (2004).

\bibitem{furhapter2005spiral}
S.~F{\"u}rhapter, A.~Jesacher, S.~Bernet, and M.~Ritsch-Marte, \enquote{Spiral
  phase contrast imaging in microscopy,} {\protect\JournalTitle{Optics
  Express}} \textbf{13}, 689--694 (2005).

\bibitem{maurer2011spatial}
C.~Maurer, A.~Jesacher, S.~Bernet, and M.~Ritsch-Marte, \enquote{What spatial
  light modulators can do for optical microscopy,} {\protect\JournalTitle{Laser
  \& Photonics Reviews}} \textbf{5}, 81--101 (2011).

\bibitem{niziev1999influence}
V.~Niziev and A.~Nesterov, \enquote{Influence of beam polarization on laser
  cutting efficiency,} {\protect\JournalTitle{Journal of Physics D: Applied
  Physics}} \textbf{32}, 1455 (1999).

\bibitem{willner2015optical}
A.~E. Willner, H.~Huang, Y.~Yan, Y.~Ren, N.~Ahmed, G.~Xie, C.~Bao, L.~Li,
  Y.~Cao, Z.~Zhao, J.~Wang, M.~P.~J. Lavery, M.~Tur, S.~Ramachandran, A.~F.
  Molisch, N.~Ashrafi, and S.~Ashrafi, \enquote{Optical communications using
  orbital angular momentum beams,} {\protect\JournalTitle{Advances in optics
  and photonics}} \textbf{7}, 66--106 (2015).

\bibitem{shen2019optical}
Y.~Shen, X.~Wang, Z.~Xie, C.~Min, X.~Fu, Q.~Liu, M.~Gong, and X.~Yuan,
  \enquote{Optical vortices 30 years on: Oam manipulation from topological
  charge to multiple singularities,} {\protect\JournalTitle{Light: Science \&
  Applications}} \textbf{8}, 1--29 (2019).

\bibitem{wang2018recent}
X.~Wang, Z.~Nie, Y.~Liang, J.~Wang, T.~Li, and B.~Jia, \enquote{Recent advances
  on optical vortex generation,} {\protect\JournalTitle{Nanophotonics}}
  \textbf{7}, 1533--1556 (2018).

\bibitem{yonezawa2006generation}
K.~Yonezawa, Y.~Kozawa, and S.~Sato, \enquote{Generation of a radially
  polarized laser beam by use of the birefringence of a c-cut nd: Yvo4
  crystal,} {\protect\JournalTitle{Optics letters}} \textbf{31}, 2151--2153
  (2006).

\bibitem{yoshida2010alignment}
H.~Yoshida, K.~Tagashira, T.~Kumagai, A.~Fujii, and M.~Ozaki,
  \enquote{Alignment-to-polarization projection in dye-doped nematic liquid
  crystal microlasers,} {\protect\JournalTitle{Optics Express}} \textbf{18},
  12562--12568 (2010).

\bibitem{Naidoo2016}
D.~Naidoo, F.~S. Roux, A.~Dudley, I.~Litvin, B.~Piccirillo, L.~Marrucci, and
  A.~Forbes, \enquote{{Controlled generation of higher-order Poincare sphere
  beams from a laser},} {\protect\JournalTitle{Nat. Photonics}} \textbf{10},
  327--332 (2016).

\bibitem{slussarenko2011tunable}
S.~Slussarenko, A.~Murauski, T.~Du, V.~Chigrinov, L.~Marrucci, and
  E.~Santamato, \enquote{Tunable liquid crystal q-plates with arbitrary
  topological charge,} {\protect\JournalTitle{Optics express}} \textbf{19},
  4085--4090 (2011).

\bibitem{humar2009electrically}
M.~Humar, M.~Ravnik, S.~Pajk, and I.~Mu{\v{s}}evi{\v{c}}, \enquote{Electrically
  tunable liquid crystal optical microresonators,}
  {\protect\JournalTitle{Nature photonics}} \textbf{3}, 595--600 (2009).

\bibitem{du2004electrically}
F.~Du, Y.-Q. Lu, and S.-T. Wu, \enquote{Electrically tunable liquid-crystal
  photonic crystal fiber,} {\protect\JournalTitle{Applied physics letters}}
  \textbf{85}, 2181--2183 (2004).

\bibitem{LcvortexREVIEW}
R.~Barboza, U.~Bortolozzo, M.~G. Clerc, S.~Residori, and E.~Vidal-Henriquez,
  \enquote{{Optical vortex induction via light--matter interaction in
  liquid-crystal media},} {\protect\JournalTitle{Adv. Opt. Photonics}}
  \textbf{7}, 635--683 (2015).

\bibitem{Brasselet2009}
E.~Brasselet, N.~Murazawa, H.~Misawa, and S.~Juodkazis, \enquote{{Optical
  vortices from liquid crystal droplets},} {\protect\JournalTitle{Phys. Rev.
  Lett.}} \textbf{103}, 4--7 (2009).

\bibitem{Loussert2013Brasselet}
C.~Loussert, U.~Delabre, and E.~Brasselet, \enquote{{Manipulating the orbital
  angular momentum of light at the micron scale with nematic disclinations in a
  liquid crystal film},} {\protect\JournalTitle{Phys. Rev. Lett.}}
  \textbf{111}, 1--4 (2013).

\bibitem{vcanvcula2014generation}
M.~{\v{C}}an{\v{c}}ula, M.~Ravnik, and S.~{\v{Z}}umer, \enquote{Generation of
  vector beams with liquid crystal disclination lines,}
  {\protect\JournalTitle{Physical Review E}} \textbf{90}, 022503 (2014).

\bibitem{donato2014polarization}
M.~G. Donato, J.~Hernandez, A.~Mazzulla, C.~Provenzano, R.~Saija, R.~Sayed,
  S.~Vasi, A.~Magazz{\`{u}}, P.~Pagliusi, R.~Bartolino, and Others,
  \enquote{{Polarization-dependent optomechanics mediated by chiral
  microresonators},} {\protect\JournalTitle{Nat. Commun.}} \textbf{5}, 3656
  (2014).

\bibitem{Smalyukh2010}
I.~I. Smalyukh, Y.~Lansac, N.~A. Clark, and R.~P. Trivedi,
  \enquote{{Three-dimensional structure and multistable optical switching of
  triple-twisted particle-like excitations in anisotropic fluids},}
  {\protect\JournalTitle{Nat Mater}} \textbf{9}, 139--145 (2010).

\bibitem{Barboza2012}
R.~Barboza, U.~Bortolozzo, G.~Assanto, E.~Vidal-Henriquez, M.~G. Clerc, and
  S.~Residori, \enquote{{Vortex Induction via Anisotropy Stabilized
  Light-Matter Interaction},} {\protect\JournalTitle{Phys. Rev. Lett.}}
  \textbf{109}, 1--5 (2012).

\bibitem{Coles2010}
H.~Coles and S.~Morris, \enquote{{Liquid-crystal lasers},}
  {\protect\JournalTitle{Nat. Photonics}} \textbf{4}, 676--685 (2010).

\bibitem{humar2011surfactant}
M.~Humar and I.~Mu{\v{s}}evi{\v{c}}, \enquote{{Surfactant sensing based on
  whispering-gallery-mode lasing in liquid-crystal microdroplets},}
  {\protect\JournalTitle{Opt. Express}} \textbf{19}, 19836--19844 (2011).

\bibitem{Nys2014}
I.~Nys, J.~Beeckman, and K.~Neyts, \enquote{{Electrically tunable Fabry–Perot
  lasing in nematic liquid crystal cells},} {\protect\JournalTitle{J. Opt. Soc.
  Am. B}} \textbf{31}, 1516 (2014).

\bibitem{papivc2021topological}
M.~Papi{\v{c}}, U.~Mur, K.~P. Zuhail, M.~Ravnik, I.~Mu{\v{s}}evi{\v{c}}, and
  M.~Humar, \enquote{Topological liquid crystal superstructures as structured
  light lasers,} {\protect\JournalTitle{Proceedings of the National Academy of
  Sciences}} \textbf{118}, e2110839118 (2021).

\bibitem{vcopar2015knot}
S.~{\v{C}}opar, U.~Tkalec, I.~Mu{\v{s}}evi{\v{c}}, and S.~{\v{Z}}umer,
  \enquote{Knot theory realizations in nematic colloids,}
  {\protect\JournalTitle{Proceedings of the National Academy of Sciences}}
  \textbf{112}, 1675--1680 (2015).

\bibitem{berteloot2020ring}
B.~Berteloot, I.~Nys, G.~Poy, J.~Beeckman, and K.~Neyts, \enquote{Ring-shaped
  liquid crystal structures through patterned planar photo-alignment,}
  {\protect\JournalTitle{Soft matter}} \textbf{16}, 4999--5008 (2020).

\bibitem{tasinkevych2014splitting}
M.~Tasinkevych, M.~G. Campbell, and I.~I. Smalyukh, \enquote{Splitting,
  linking, knotting, and solitonic escape of topological defects in nematic
  drops with handles,} {\protect\JournalTitle{Proceedings of the National
  Academy of Sciences}} \textbf{111}, 16268--16273 (2014).

\bibitem{machon2014knotted}
T.~Machon and G.~P. Alexander, \enquote{Knotted defects in nematic liquid
  crystals,} {\protect\JournalTitle{Physical review letters}} \textbf{113},
  027801 (2014).

\bibitem{martinez2014mutually}
A.~Martinez, M.~Ravnik, B.~Lucero, R.~Visvanathan, S.~{\v{Z}}umer, and I.~I.
  Smalyukh, \enquote{Mutually tangled colloidal knots and induced defect loops
  in nematic fields,} {\protect\JournalTitle{Nature materials}} \textbf{13},
  258--263 (2014).

\bibitem{sevc2012geometrical}
D.~Se{\v{c}}, T.~Porenta, M.~Ravnik, and S.~{\v{Z}}umer, \enquote{Geometrical
  frustration of chiral ordering in cholesteric droplets,}
  {\protect\JournalTitle{Soft Matter}} \textbf{8}, 11982--11988 (2012).

\bibitem{posnjak2016points}
G.~Posnjak, S.~{\v{C}}opar, and I.~Mu{\v{s}}evi{\v{c}}, \enquote{Points,
  skyrmions and torons in chiral nematic droplets,}
  {\protect\JournalTitle{Scientific reports}} \textbf{6}, 26361 (2016).

\bibitem{cryst12010094}
Y.~Shen and I.~Dierking, \enquote{Recent progresses on experimental
  investigations of topological and dissipative solitons in liquid crystals,}
  {\protect\JournalTitle{Crystals}} \textbf{12}, 94 (2022).

\bibitem{hess2020control}
A.~J. Hess, G.~Poy, J.-S.~B. Tai, S.~{\v{Z}}umer, and I.~I. Smalyukh,
  \enquote{Control of light by topological solitons in soft chiral birefringent
  media,} {\protect\JournalTitle{Physical Review X}} \textbf{10}, 031042
  (2020).

\bibitem{varanytsia2017topology}
A.~Varanytsia, G.~Posnjak, U.~Mur, V.~Joshi, K.~Darrah, I.~Mu{\v{s}}evi{\v{c}},
  S.~{\v{C}}opar, and L.-C. Chien, \enquote{Topology-commanded optical
  properties of bistable electric-field-induced torons in cholesteric bubble
  domains,} {\protect\JournalTitle{Scientific reports}} \textbf{7}, 1--8
  (2017).

\bibitem{ackerman2017diversity}
P.~J. Ackerman and I.~I. Smalyukh, \enquote{Diversity of knot solitons in
  liquid crystals manifested by linking of preimages in torons and hopfions,}
  {\protect\JournalTitle{Physical Review X}} \textbf{7}, 011006 (2017).

\bibitem{nych2017spontaneous}
A.~Nych, J.-i. Fukuda, U.~Ognysta, S.~{\v{Z}}umer, and I.~Mu{\v{s}}evi{\v{c}},
  \enquote{Spontaneous formation and dynamics of half-skyrmions in a chiral
  liquid-crystal film,} {\protect\JournalTitle{Nature Physics}} \textbf{13},
  1215--1220 (2017).

\bibitem{nys2020patterned}
I.~Nys, \enquote{Patterned surface alignment to create complex
  three-dimensional nematic and chiral nematic liquid crystal structures,}
  {\protect\JournalTitle{Liquid Crystals Today}} \textbf{29}, 65--83 (2020).

\bibitem{tartan2017generation}
C.~Tartan, P.~Salter, T.~Wilkinson, M.~Booth, S.~Morris, and S.~Elston,
  \enquote{Generation of 3-dimensional polymer structures in liquid crystalline
  devices using direct laser writing,} {\protect\JournalTitle{RSC advances}}
  \textbf{7}, 507--511 (2017).

\bibitem{he2019novel}
Z.~He, G.~Tan, D.~Chanda, and S.-T. Wu, \enquote{Novel liquid crystal photonic
  devices enabled by two-photon polymerization,} {\protect\JournalTitle{Optics
  express}} \textbf{27}, 11472--11491 (2019).

\bibitem{deGennesPG_1993}
P.~G. de~Gennes and J.~Prost, \emph{Physics of Liquid Crystals} (Clarendon
  Press, Oxford, 1993).

\bibitem{kleman2006topological}
M.~Kleman and O.~D. Lavrentovich, \enquote{Topological point defects in nematic
  liquid crystals,} {\protect\JournalTitle{Philosophical Magazine}}
  \textbf{86}, 4117--4137 (2006).

\bibitem{cardano2012polarization}
F.~Cardano, E.~Karimi, S.~Slussarenko, L.~Marrucci, C.~de~Lisio, and
  E.~Santamato, \enquote{Polarization pattern of vector vortex beams generated
  by q-plates with different topological charges,}
  {\protect\JournalTitle{Applied optics}} \textbf{51}, C1--C6 (2012).

\bibitem{ackerman2012optical}
P.~J. Ackerman, Z.~Qi, and I.~I. Smalyukh, \enquote{Optical generation of
  crystalline, quasicrystalline, and arbitrary arrays of torons in confined
  cholesteric liquid crystals for patterning of optical vortices in laser
  beams,} {\protect\JournalTitle{Physical Review E}} \textbf{86}, 021703
  (2012).

\bibitem{barboza2013harnessing}
R.~Barboza, U.~Bortolozzo, G.~Assanto, E.~Vidal-Henriquez, M.~G. Clerc, and
  S.~Residori, \enquote{Harnessing optical vortex lattices in nematic liquid
  crystals,} {\protect\JournalTitle{Physical review letters}} \textbf{111},
  093902 (2013).

\bibitem{gray1973new}
G.~W. Gray, K.~J. Harrison, and J.~Nash, \enquote{New family of nematic liquid
  crystals for displays,} {\protect\JournalTitle{Electronics Letters}}
  \textbf{9}, 130--131 (1973).

\bibitem{sharma2021electronic}
D.~Sharma, G.~Tiwari, and S.~N. Tiwari, \enquote{Electronic and electro-optical
  properties of 5cb and 5ct liquid crystal molecules: A comparative dft study,}
  {\protect\JournalTitle{Pramana}} \textbf{95}, 1--7 (2021).

\bibitem{wu2019high}
X.~Wu, Y.~Wang, Q.~Chen, Y.-C. Chen, X.~Li, L.~Tong, and X.~Fan,
  \enquote{High-q, low-mode-volume microsphere-integrated fabry--perot cavity
  for optofluidic lasing applications,} {\protect\JournalTitle{Photonics
  Research}} \textbf{7}, 50--60 (2019).

\bibitem{rosales2018review}
C.~Rosales-Guzm{\'a}n, B.~Ndagano, and A.~Forbes, \enquote{A review of complex
  vector light fields and their applications,} {\protect\JournalTitle{Journal
  of Optics}} \textbf{20}, 123001 (2018).

\bibitem{snyder2012optical}
A.~W. Snyder and J.~Love, \emph{Optical waveguide theory} (Springer Science \&
  Business Media, 2012).

\bibitem{hall1996vector}
D.~G. Hall, \enquote{Vector-beam solutions of maxwell’s wave equation,}
  {\protect\JournalTitle{Optics letters}} \textbf{21}, 9--11 (1996).

\bibitem{zaoui2014bridging}
W.~S. Zaoui, A.~Kunze, W.~Vogel, M.~Berroth, J.~Butschke, F.~Letzkus, and
  J.~Burghartz, \enquote{Bridging the gap between optical fibers and silicon
  photonic integrated circuits,} {\protect\JournalTitle{Optics express}}
  \textbf{22}, 1277--1286 (2014).

\bibitem{tovar1998production}
A.~A. Tovar, \enquote{Production and propagation of cylindrically polarized
  laguerre--gaussian laser beams,} {\protect\JournalTitle{JOSA A}} \textbf{15},
  2705--2711 (1998).

\bibitem{greene1998properties}
P.~L. Greene and D.~G. Hall, \enquote{Properties and diffraction of vector
  bessel--gauss beams,} {\protect\JournalTitle{JOSA A}} \textbf{15}, 3020--3027
  (1998).

\bibitem{cao2002lasing}
W.~Cao, A.~Munoz, P.~Palffy-Muhoray, and B.~Taheri, \enquote{Lasing in a
  three-dimensional photonic crystal of the liquid crystal blue phase ii,}
  {\protect\JournalTitle{Nature materials}} \textbf{1}, 111--113 (2002).

\bibitem{mysliwiec2021liquid}
J.~Mysliwiec, A.~Szukalska, A.~Szukalski, and L.~Sznitko, \enquote{Liquid
  crystal lasers: the last decade and the future,}
  {\protect\JournalTitle{Nanophotonics}}  (2021).

\bibitem{kolle2018progress}
M.~Kolle and S.~Lee, \enquote{Progress and opportunities in soft photonics and
  biologically inspired optics,} {\protect\JournalTitle{Advanced Materials}}
  \textbf{30}, 1702669 (2018).

\bibitem{wu2016optical}
L.~Wu, J.~He, W.~Shang, T.~Deng, J.~Gu, H.~Su, Q.~Liu, W.~Zhang, and D.~Zhang,
  \enquote{Optical functional materials inspired by biology,}
  {\protect\JournalTitle{Advanced Optical Materials}} \textbf{4}, 195--224
  (2016).

\bibitem{humar2017toward}
M.~Humar, S.~J. Kwok, M.~Choi, A.~K. Yetisen, S.~Cho, and S.-H. Yun,
  \enquote{Toward biomaterial-based implantable photonic devices,}
  {\protect\JournalTitle{Nanophotonics}} \textbf{6}, 414--434 (2017).

\bibitem{kim2011self}
S.-H. Kim, S.~Y. Lee, S.-M. Yang, and G.-R. Yi, \enquote{Self-assembled
  colloidal structures for photonics,} {\protect\JournalTitle{NPG Asia
  Materials}} \textbf{3}, 25--33 (2011).

\bibitem{zhuo2019hierarchical}
M.-P. Zhuo, J.-J. Wu, X.-D. Wang, Y.-C. Tao, Y.~Yuan, and L.-S. Liao,
  \enquote{Hierarchical self-assembly of organic heterostructure nanowires,}
  {\protect\JournalTitle{Nature communications}} \textbf{10}, 1--9 (2019).

\bibitem{fan2010self}
J.~A. Fan, C.~Wu, K.~Bao, J.~Bao, R.~Bardhan, N.~J. Halas, V.~N. Manoharan,
  P.~Nordlander, G.~Shvets, and F.~Capasso, \enquote{Self-assembled plasmonic
  nanoparticle clusters,} {\protect\JournalTitle{science}} \textbf{328},
  1135--1138 (2010).

\bibitem{chen2020soft}
Y.~Chen, B.~Ai, and Z.~J. Wong, \enquote{Soft optical metamaterials,}
  {\protect\JournalTitle{Nano Convergence}} \textbf{7}, 1--17 (2020).

\bibitem{humar2015intracellular}
M.~Humar and S.~H. Yun, \enquote{Intracellular microlasers,}
  {\protect\JournalTitle{Nature photonics}} \textbf{9}, 572--576 (2015).

\bibitem{demchenko2020photobleaching}
A.~P. Demchenko, \enquote{Photobleaching of organic fluorophores: quantitative
  characterization, mechanisms, protection,} {\protect\JournalTitle{Methods and
  applications in fluorescence}} \textbf{8}, 022001 (2020).

\bibitem{van2018grand}
J.~van~der Gucht, \enquote{Grand challenges in soft matter physics,}
  {\protect\JournalTitle{Frontiers in Physics}} \textbf{6}, 87 (2018).

\bibitem{zhang2015organic}
C.~Zhang, C.-L. Zou, Y.~Zhao, C.-H. Dong, C.~Wei, H.~Wang, Y.~Liu, G.-C. Guo,
  J.~Yao, and Y.~S. Zhao, \enquote{Organic printed photonics: From microring
  lasers to integrated circuits,} {\protect\JournalTitle{Science Advances}}
  \textbf{1}, e1500257 (2015).

\bibitem{lu2017soft}
T.~Lu, E.~J. Markvicka, Y.~Jin, and C.~Majidi, \enquote{Soft-matter printed
  circuit board with uv laser micropatterning,} {\protect\JournalTitle{ACS
  applied materials \& interfaces}} \textbf{9}, 22055--22062 (2017).

\bibitem{fu2012all}
Y.~Fu, X.~Hu, C.~Lu, S.~Yue, H.~Yang, and Q.~Gong, \enquote{All-optical logic
  gates based on nanoscale plasmonic slot waveguides,}
  {\protect\JournalTitle{Nano letters}} \textbf{12}, 5784--5790 (2012).

\bibitem{salmanpour2015photonic}
A.~Salmanpour, S.~Mohammadnejad, and A.~Bahrami, \enquote{Photonic crystal
  logic gates: an overview,} {\protect\JournalTitle{Optical and Quantum
  Electronics}} \textbf{47}, 2249--2275 (2015).

\end{thebibliography}
\end{document}